\documentstyle[preprint,aps,epsfig,floats]{revtex}
\def\journal#1, #2, 1#3#4#5#6, #7    {
    {\sl #1~}{\bf #2}, #7    (1#3#4#5)#6}
\def\pr{\journal Phys. Rev., }

\def\prl{\journal Phys. Rev. Lett., }

\def\np{\journal Nucl. Phys., }
\def\pl{\journal Phys. Lett., }

\def\mpl{\journal Mod. Phys. Lett., }

\def\jmp{\journal J. Math. Phys., }
\def\jp{\journal J. Phys., }
\newcommand{\beq}[1]{\begin{equation}\label{#1}}
\newcommand\eeq{\end{equation}}
\newcommand{\ba}[1]{\begin{eqnarray}\label{#1}}
\newcommand{\baa}{\begin{eqnarray}}
\newcommand\ea{\end{eqnarray}}
\newcommand{\bee}{\begin{equation}}
\def\nn{\nonumber \\}
\def\l{\lambda}

\newcommand{\cl}{Calogero}
\newcommand{\h}{Hamiltonian}
\newcommand{\col}{collective}

\newcommand{\B}[1]{{\bf #1}}
\newcommand{\gsim}{\stackrel{\rm >} {\scriptstyle \sim}}
\newcommand{\lsim}{\stackrel{\rm <} {\scriptstyle \sim}}

\begin{document}
\draft
\title{Quantum fluctuations of the 
  Chern-Simons theory \\ and  dynamical dimensional reduction}
\author{Ivan Andri\'c, Velimir Bardek, and Larisa Jonke
\footnote{e-mail address: 
andric@thphys.irb.hr \\ \hspace*{3cm} bardek@thphys.irb.hr \\ \hspace*{3cm} 
larisa@thphys.irb.hr }}
\address{Theoretical Physics Division,\\
Rudjer Bo\v skovi\'c Institute, P.O. Box 1016,\\
10001 Zagreb, CROATIA}
\maketitle
\begin{abstract}
We consider a large-N Chern-Simons theory for the attractive bosonic 
matter (Jackiw-Pi model) in the \h \ \col-field 
approach based on the 1/N expansion. We show that the dynamics of 
low-lying density 
excitations around the ground-state vortex configuration is 
equivalent to that of  the Sutherland model. 
The relationship between 
the Chern-Simons coupling constant $\l$ and the 
\cl-Sutherland statistical parameter $\l_s$ 
signalizes some sort of statistical transmutation accompanying the 
dimensional reduction of the initial problem.
\end{abstract}

PACS number(s): 11.10.Lm 74.20.Kk 03.65.Sq 05.30.-d 
\vspace{1cm}

Gauge models of a scalar field with the Chern-Simons term\cite{100} in $2+1$
space-time dimensions are known to support soliton or vortex solutions\cite
{111,222}. By using the nonrelativistic field theory of the self-attracted
bosonic matter minimally coupled to an Abelian Chern-Simons gauge field, the
authors of Ref.\cite{111} have shown that there exists a static self-dual
soliton solution for a specific choice of the coupling constant. We have
rederived this soliton solution in the \col-field approach by including
higher-order terms in the $1/N$ expansion\cite{4}. In our approach, this
soliton solution saturates the Bogomol'nyi bound and does not receive quantum
corrections to its energy in the next-to-leading approximation.

In this paper we analyze the quantum dynamics of low-lying density 
fluctuations around a specific vortex solution and show that it is 
equivalent to the dynamics of quantum fluctuations in  the Calogero-Sutherland 
model\cite{cals}.
There exist a number of papers\cite{all} that elucidate the connection 
between the Chern-Simons-based anyonic physics in the fractional quantum
Hall effect and the Calogero-Sutherland model, but it should be emphasized 
that we are working in a completely different physical situation.
We are trying to establish a dynamical reduction of the Jackiw-Pi model to the 
Calogero-Sutherland one.
In Ref.\cite{77} we conjectured the form of  quantum fluctuations in the 
Jackiw-Pi model and that allowed us to identify the dynamics of these 
fluctuations with those of the Calogero-Sutherland model. In this paper, we are 
looking for the same result  using a different approach and thus 
indirectly confirming the conjecture made in \cite{77}.

We begin our analysis of the Jackiw-Pi model by 
repeating the main results of 
Ref.\cite{4}. The \col -field approach to the model is described by the 
\h 
\bee \label{1'}
H= \frac{1}{2}\int d^2\B r\rho(\B r)\left[ \B\nabla\pi(\B r)+\hat{n}
\times\left(\frac{1}{2}\frac
{\B\nabla\rho(\B r)}{\rho(\B r)}+|\l|\int d^2\B r'\rho(\B r')
\frac{\B r-\B r'}{|\B r-\B r'|^2}-v\frac{\B r-\B R}{|\B r-\B R|^2}
\right)\right]^2, \eeq
where $\hat{n}$ is the unit vector perpendicular to the plane in which 
particles move, the dimensionless constant $\l$ is the so-called 
statistical parameter, 
the vorticity $v$ is a dimensionless integer, and $\pi(\B r)$ is the 
canonical conjugate of the \col \ field 
$\rho(\B r)$: \bee \label{com}
[\nabla\pi(\B r),\rho(\B r')]=-i\nabla\delta(\B r-\B r') .\eeq
The leading part of the \col -field \h \ in the $1/N$ expansion is 
given by the effective potential
\bee \label{veff}
V_{\rm eff}=\frac{1}{2}\int d^2\B r\rho(\B r)\left(\frac{1}{2}\frac
{\B\nabla\rho(\B r)}{\rho(\B r)}+|\l|\int d^2\B r'\rho(\B r')
\frac{\B r-\B r'}{|\B r-\B r'|^2}-v\frac{\B r-\B R}{|\B r-\B R|^2}
\right)^2. \eeq
Owing to the positive definiteness of the effective potential (\ref{veff}), 
the Bogomol'nyi limit appears. The Bogomol'nyi bound is saturated by the 
 positive normalizable solution $\rho_0(\B r)$ of the  Liouville-type equation
\bee \label{lx} 
\Delta\ln\rho_0(\B r)+4|\l|\pi\rho_0(\B r)=4\pi v\delta(\B r) . \eeq
It has been shown in\cite{4} that there exists a radially symmetric, 
positive, and normalizable \col -field configuration that minimizes the 
energy (\ref{1'}). It is given by the vortex form
\beq y
\rho_0(r)=\frac{|\l| N^2}{2\pi r^2}\left[\left(\frac{r_0}{r}\right)^{
\frac{N|\l|}{2}}+\left(\frac{r}{r_0}\right)^{\frac{N|\l|}{2}}\right]^{-2} .
\eeq
The vorticity $v$ is fixed by the normalization condition and is given by
\bee v=N\frac{|\l|}{2}-1 .\eeq
The parameter $r_0$ reflects the scale invariance of the problem and cannot be
determined. Now, if N is large enough, we can replace the soliton 
configuration $\rho_0(r)$ by the $\delta$ profile:
\beq n \rho_0(r)=\frac{N}{2\pi}\frac{\delta(r-r_0)}{r_0} .\eeq
Here we have used the well-known representation of the $\delta$-function:
\bee \delta(x)=\lim\limits_{\epsilon\to 0}\frac{{\rm exp}(x/\epsilon)}
{\epsilon[1+
{\rm exp}(x/\epsilon)]^2}\;,\;\epsilon=\frac{1}{N|\l|} .\eeq
At this point we
analyze the dynamics of the collective-field  excitations around the 
ground-state
solution of the Jackiw-Pi  model. We perform the $1/N$ expansion
of the \col \ field $\rho(\B r)$ in the form
\bee\label{exp1}
\rho(\B r)=\rho_0(\B r)+\eta(\B r) , \eeq
where $\rho_0(\B r)$ is the ground-state semiclassical configuration and
$\eta(\B r)$ a small density quantum fluctuation around  $\rho_0(\B r)$.
Inserting (\ref{exp1}) in (\ref{1'}) and expanding in $\eta(\B r)$, we obtain
the leading term $V_{\rm eff}(\rho_0)$ and the \h \ 
quadratic in fluctuations and
its canonical conjugate. After introducing the operators
\baa \label{AA}
A(z)=\frac{\partial\pi}{\partial z}-i\frac{\partial}{\partial z}\left(
\frac{\eta}{2\rho_0}+|\l|\int d^2\B r'\ln |z-z'|\eta(z')\right), \nn
A^{\dagger}(z)=\frac{\partial\pi}{\partial\bar z}
+i\frac{\partial}{\partial\bar z}\left(
\frac{\eta}{2\rho_0}+|\l|\int d^2\B r'\ln |z-z'|\eta(z')\right), \ea
with the c-number commutator (all other vanishing)
\bee\label{comA}
\left[A(z),A^{\dagger}(z')\right]=2\frac{\partial^2}{\partial z\partial\bar z'}
\left(\frac{\delta^2(z-z')}{2\rho_0}+|\l|\ln |z-z'|\right), \eeq
we are left with a \h \ that governs the dynamics of low-lying excitations
in the form:
\bee\label{H2}
H=2\int d^2\B r\rho_0(z)A^{\dagger}(z)A(z) .\eeq
There are some subtleties involving the ordering of the operators $A$ and 
$A^{\dagger}$, but an interested reader can find all details in 
Ref. \cite{4}.

Now, to find the spectrum of low-lying excitations, 
we have to diagonalize the
\h \ (\ref{H2}).
We expand the operators $A$ and $A^{\dagger}$ in terms of a new,
complete  set of operators
\bee \label{expa}
A(z)=\sum_{\B n}\phi_{\B n}(z) a_{\B n}\;,\;
A^{\dagger}(z)=\sum_{\B n}\phi_{\B n}^*(z) a_{\B n}^{\dagger} \eeq
that satisfy the standard bosonic commmutation relations
\bee \label{coma}[a_{\B n},a_{\B m}^{\dagger}]=\delta_{\B n,\B m}\;,\;
[a_{\B n},a_{\B m}]=
 [a_{\B n}^{\dagger},a_{\B m}^{\dagger}]=0.
\eeq
We demand that the \h \ (\ref{H2}) should take the diagonal form
\bee\label{dia}
H=\sum_{\B n}\omega_{\B n}a_{\B n}^{\dagger}a_{\B n} .\eeq
Here, $\B n$ represents a pair of quantum numbers, and it is assumed that 
the sum is replaced by an integral in the  nondiscrete case.
We insert expansion (\ref{expa}) in the commutator (\ref{comA}) and apply
(\ref{coma}) to obtain the completeness relation:
\bee\label{compl}
\sum_{\B n}\phi_{\B n}(z)\phi_{\B n}^*(z')=2\frac{\partial^2}
{\partial z\partial\bar z'}
\left(\frac{\delta^2(z-z')}{2\rho_0}+|\l|\ln |z-z'|\right). \eeq
Inserting (\ref{expa}) in (\ref{H2}), and demanding (\ref{dia}),
we obtain
\bee\label{XZ}
2\int d^2\B r \rho_0(\B r)\phi_{\B m}^*(z)\phi_{\B n}(z)
=\omega_{\B m}\delta_{\B n,\B m} .\eeq
Next, we multiply relation (\ref{XZ}) by $\phi_{\B m}(z')$ and sum over 
$\B m$, apply the completeness relation (\ref{compl}), and finally 
 we obtain the equation for the
functions $\phi_{\B n}(z)$:
\bee\label{last}
\frac{1}{2}\omega_{\B n}\phi_{\B n}(z)=-\partial_z\partial_{\bar z}
\phi_{\B n}(z)
-\partial_{\bar z}
\ln\rho_0(z)\partial_z\phi_{\B n}(z)
-(\partial_z\partial_{\bar z}
\ln\rho_0(z))\phi_{\B n}(z)-\l\pi\rho_0(z)\phi_{\B n}(z). \eeq
Since we are interested in the excitations around a specific ground state, 
namely, the one given in  (\ref{n}), we use Eq. (\ref{lx}) to rewrite the last 
two terms in Eq.(\ref{last}) 
and  look for the solution
in  terms of a new function $\psi_{\B n}(r,\varphi)=\phi_{\B n}(r,\varphi)
\sqrt{\rho_0(r)}$:
\bee \label{eq}
\omega_{\B n}\psi_{\B n}(r,\varphi)=
-\frac{1}{2}\left(\B\nabla-i\B A\right)^2\psi_{\B n}(r,\varphi)
-\frac{1}{2}\left(v\delta(\B r)+\l N\frac{\delta(r-r_0)}{r_0}\right)
\psi_{\B n}(r,\varphi).
\eeq
Equation (\ref{eq}) can be interpreted as a Schr\"odinger equation for
the bosonic particle in  the magnetic field and in the additional
delta-function  potential, where
\bee\label{ab}
A_r(r,\varphi)=0\;,\;A_{\varphi}(r,\varphi)=-\frac{N\l}{2r}\;{\rm sign}
(r-r_0) \;,\;
B=\nabla\times\B A=-\frac{N\l}{r}\;\delta(r-r_0).\eeq
Notice that we can omit the  'vorticity' term in Eq. (\ref{eq}), because 
for $r\to 0$ the wave function must vanish at least as $r^{2}$. 
The operator on the right-hand  side of Eq. (\ref{eq}) commutes
 with the angular 
momentum operator ($-i\partial /\partial\varphi $), so 
we extract a factor $\exp(\pm in\varphi)$ from the eigenfunction, and obtain:
\bee \label{123}
\frac{1}{r}\partial_r(r\partial_r)\tilde\psi_0(r)+\left[ 2\omega_n
-\frac{1}{r^2}
( \frac{N\l}{2}{\rm sign}(r-r_0)\pm n)^2\right]\tilde\psi_0(r)
+\frac{N\l}{r_0}
\delta(r-r_0)\tilde\psi_0(r)=0, \eeq
where $\tilde\psi_0(r)$ represents the
lowest-energy solution.
We have two classes of solution. The first one is
\bee\label{f1}
\psi_n(r,\varphi)=\exp(\pm in\varphi)\times\left\{
\begin{array}{ll}
  J_{\nu}(x)  & {\rm for}\;\; r<r_0, \\
  N_{\mu}(x)  & {\rm for}\;\; r>r_0,
\end{array} \right. \eeq
and the second one is
\bee\label{f2}
\psi_n(r,\varphi)=\left\{
\begin{array}{ll}
\exp(\pm in\varphi) J_{\nu}(x)  & {\rm for}\;\; r<r_0 ,\\
\exp(\mp in\varphi) N_{\nu}(x)  & {\rm for}\;\; r>r_0,
\end{array} \right. \eeq
where $J_{\nu}(x)$ and $ N_{\nu}(x)$ are 
Bessel functions, $x={\sqrt{2\omega_n}}\;r$,
and $\mu(\nu)=N\l/2\pm n\;(N\l/2\mp n)$.
%
\begin{figure}
\centerline{ \epsfig{file=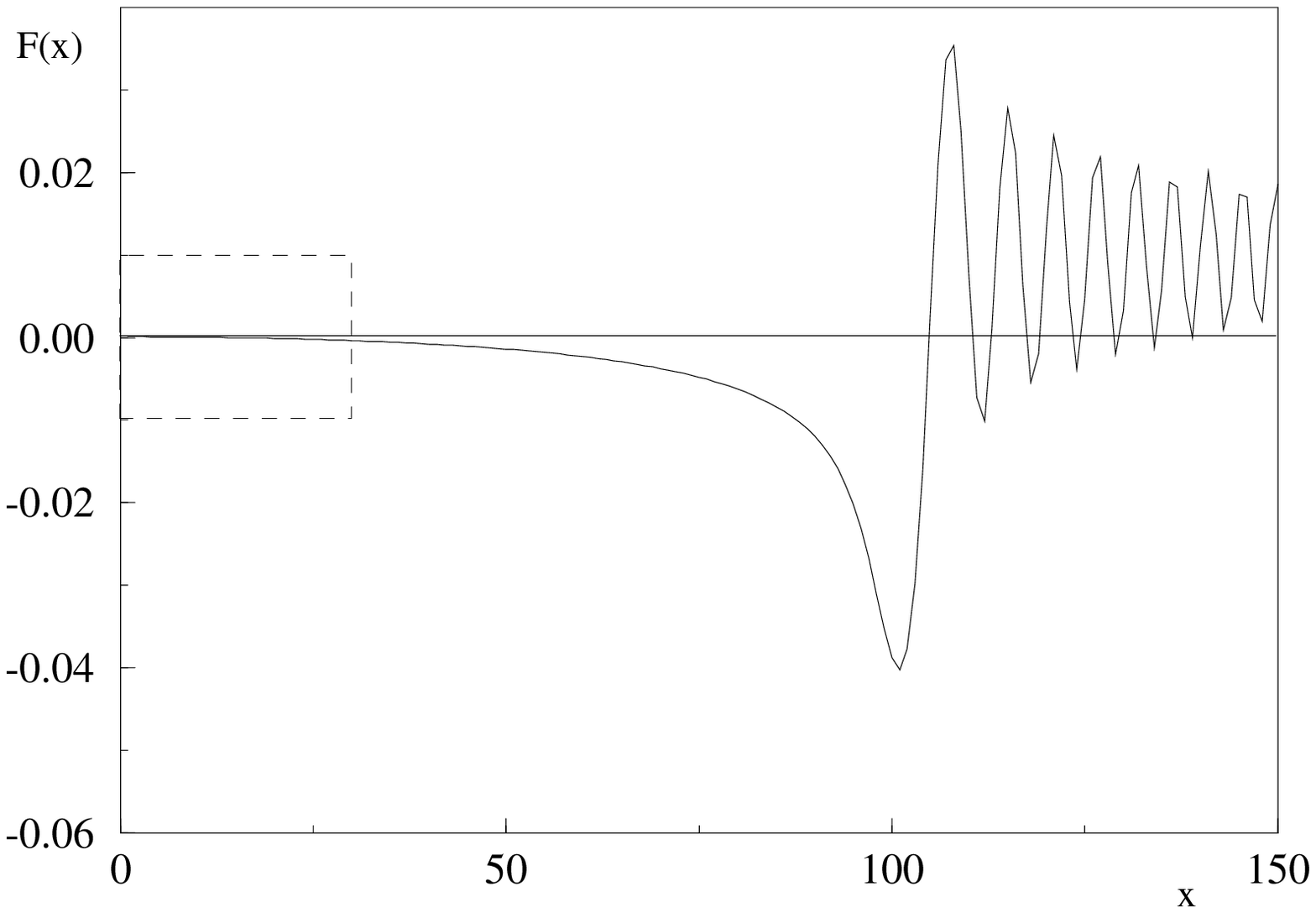,height=5cm,width=6cm,silent=}
 \hspace{2cm}\epsfig{file=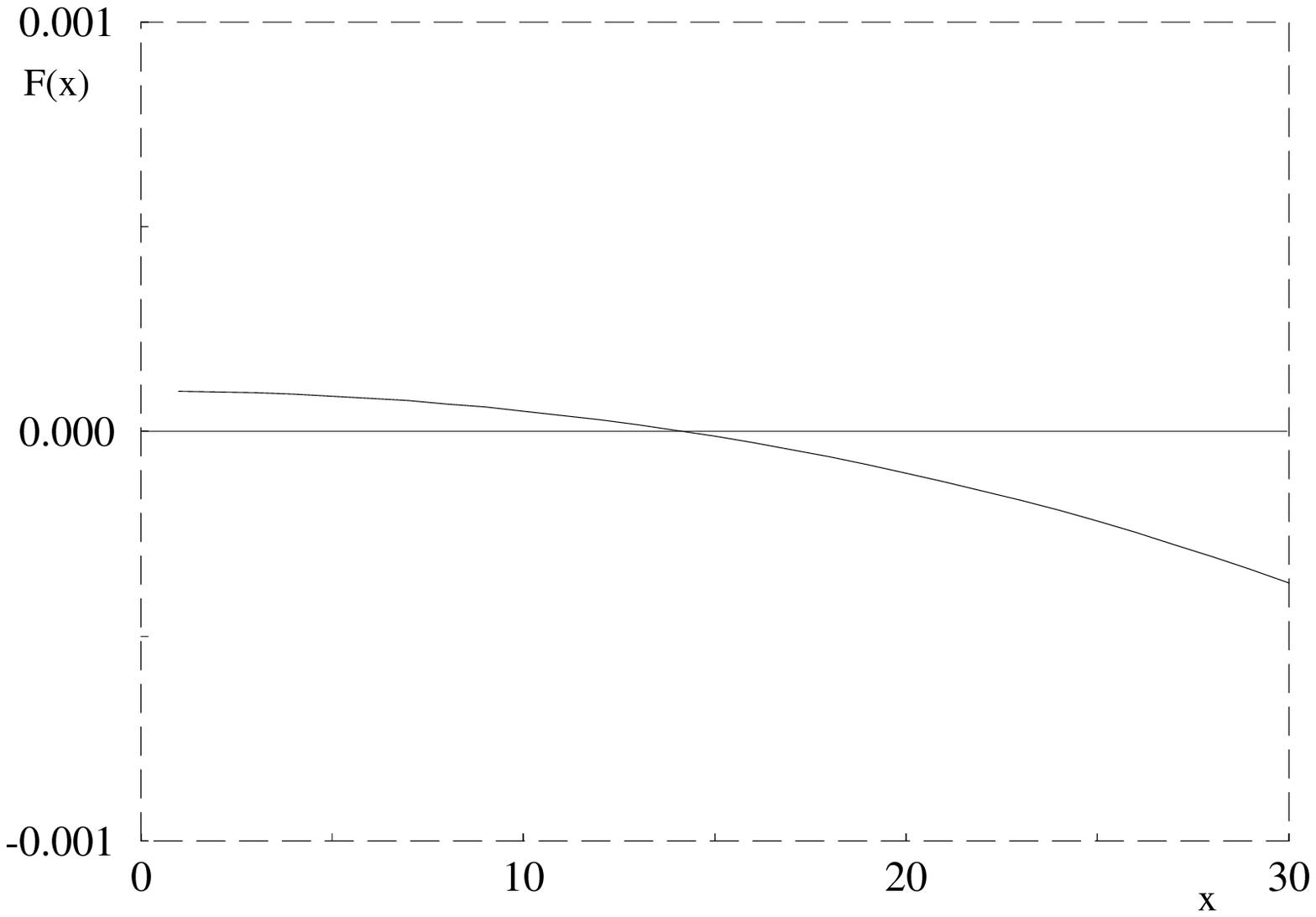,height=5cm,width=6cm,silent=}}
 \caption{On the left, the function $F(x)$ 
 is plotted for $\nu=\mu=101$, 
 $N\l/2=100$. The area in the small box is shown enlarged on the right.}
 \label{fig1}
\end{figure}
%
To find out the energy of the solutions (\ref{f1}) and (\ref{f2}), we have 
to solve the equation of the type
\bee \label{Bess}
F(x)\equiv J_{\nu}(x)N_{\nu(\mu)}(x)-\frac{2}{N\l\pi}=0 ,\eeq
using the  expansion for Bessel function for large $\nu(\mu)$\cite{table}. 
For  small $x$, 
 i. e., for low-energy fluctuations, 
the function $F(x)$ 
decreases monotonously and has only one zero, and this is precisely what we are 
looking for. We illustrate the behavior of the function F(x) in
Fig. \ref{fig1}. Notice that the next zero of the function $F(x)$ is at 
$x\sim\nu$. This  means that $\omega_n\gsim N^2/r_0^2$, and  it costs 
infinite amount of energy to reach the next state. Effectively, our
lowest-energy state is the only relevant state in large $N$.
Solving Eq.(\ref{Bess}) for $\mu\neq\nu$, 
we obtain the zero mode $(\omega_n=0)$, 
but for $\mu=\nu$ we 
have $2\omega_nr_0^2=4n^2\mp nN\l$. Since $\omega_n$ is positive by definition,
 we finally have
\bee\label{f3}
\psi_n(r,\varphi)=\left\{
\begin{array}{ll}
\exp(-i|n|\varphi) J_{\nu}(x)  & {\rm for}\;\; r<r_0 ,\\
\exp(+i|n|\varphi) N_{\nu}(x)  & {\rm for}\;\; r>r_0,
\end{array} \right. \eeq
where $\nu=\frac{1}{2}N\l+|n|$ and 
\bee \label{disp}\omega_n=\frac{1}{r_0^2}(2n^2+|n|\frac{N\l}{2}). \eeq
We can compare the above relation with the dispersion relation of 
fluctuations around the constant solution  in the
Sutherland model on a circle of radius $r_0$\cite{unp,mp}:
\bee \label{disp2}\omega_n^s=\frac{1}{r_0^2}\left(\frac{1-\l_s}{2}n^2+
\frac{N_s\l_s}{4}|n|\right). \eeq 
In order to establish a full correspondence, we should rescale the dispersion 
relation (\ref{disp}) as $\omega_n\to 4\omega_n/(1-\l_s)$, and demand a 
sort of statistical transmutation given by 
\bee\label{trans}
N\l=\frac{2 N_s\l_s}{1-\l_s}.\eeq
It is interesting to note that the above relation is invariant under the 
duality 
transformation $\l_s\to 1/\l_s$, with $N_s\to -\l_s N_s$, 
reflecting the well-known symmetry of the Sutherland model\cite{mp}.

We will now show how our result can be extended to more general vortex 
solution then that which is concentrated at the origin.
The ground-state  solution of Eq. (\ref{lx}) 
is given in the terms of the analytic function
$f(z)$:
\bee \label{r0}
\rho_0(z,\bar z)=\frac{2}{\l\pi} \frac{|f'(z)|^2}{[1+|f(z)|^2]^2}=
\frac{1}{2\l\pi}\left|\frac{f'}{
f}\right|\frac{1}{\cosh^2\ln|f|}\; .\eeq
Let us investigate configurations of vortices positioned at the origin and 
around it at the locations $z_i$ such that outside the circle of radius $R$ 
there are no vortices ($|z_i|<R$). Then $f$ is a polynomial in $z/R$ the 
degree of which is determined by the normalization condition.
 At the origin, $f$ goes like $(z/R)^{\alpha N},\;\alpha\lsim 1$.
The vortices are positioned at the zeros of $f'$. We have a {\it strong} 
vortex at $z=0$, and other vortices are inside the circle.
Owing to the normalization condition 
\baa \label{dod1}
N&=&\int dxdy\rho(z,\bar z)=\frac{i}{2\l\pi}\int_{\bf R^2} d\omega_1
=\frac{i}{2\l\pi}\int_{\partial\bf R^2}\omega_1= \nn
&=&\frac{i}{2\l\pi}\int_{\partial\bf R^2}
\left(\frac{fd\bar f}{1+f\bar f}-\frac{df \bar f}{1+f\bar f}\right) , \ea
$\ln|f|$ is proportional to $N$. 
Now, $\cosh^{-2}\ln|f|$ and therefore $\rho_0(z,\bar z)$ 
is strongly peaked at 
$|f|=1$, up to ${\cal O}(1/N^2)$, with the width $\Delta\propto 1/N$. 
The condition  $|f(z/R)|=1$ describes a closed string which in 
 the large-$N$ limit approaches the circle of radius $R$.
For illustration, take a simple example
\bee \label{example}
f(z)=\left(\frac{z}{R}\right)^n\left(\left(\frac{z}{R}\right)^m-1\right),\eeq
where  $n$, $m$ are of order $N$, $n+m=N$.
This describes the {\it strong} vortex at the origin and a certain number of 
equidistant vortices positioned on the circle around the origin. 
$\rho_0(z,\bar z)$ has a maximum on the closed string determined from $|f|=1$:
\bee\label{equ.}
\left(\frac{r}{R}\right)^{2(n+m)}-2\cos(n+m)\varphi \left(\frac{r}{R}\right)^{
n+m}+\left(\frac{r}{R}\right)^{2n}=1 .\eeq
In the large-$N$ limit, the equation $r=r(\varphi)$ is approaching a circle 
of radius $r=R+{\cal O}(1/N)$.
Therefore in the large-$N$ limit, we have a bulk-to-edge dimensional 
reduction:
$\lim\limits_{N\to \infty} \rho_0(z,\bar z)
\to\delta(|f|-1)\to \delta(r-R).$ 
 
In conclusion, we can say that in the large-$N$ limit the dynamics of
low-lying density
excitations around the vortex configuration in the Jackiw-Pi 
model is
equivalent to that of  the Sutherland model. 
Further study is still needed to fully understand the physical meaning
of this dimensional reduction  and the
statistical transmutation associated with it.

{\bf Acknowledgment}

This work was supported by the Ministry of Science and Technology of the
 Republic of Croatia under Contract No. 00980103.

\end{document}